\documentclass[a4paper]{JHEP3}

\usepackage{amssymb}
\usepackage{latexsym}

\newcommand{\cosech}{\,{\rm cosech}\,}

\newcommand{\tr}{\,{\rm tr}\,}
\newcommand{\End}{\,{\rm End}\,}
\newcommand{\diag}{\,{\rm diag}\,}
\newcommand{\Mat}{\,{\rm Mat}\,}

\newcommand{\sgn}{\,{\rm sgn}\,}

\author{James Lucietti\\  DAMTP, Centre of Mathematical
Sciences, Cambridge University, Wilberforce Road, Cambridge CB3 0WA,
UK \\ E-mail: \email{J.Lucietti@damtp.cam.ac.uk }}

\title{Canonical quantization of a massive particle on $AdS_3$} 

\maketitle

\abstract{The classical theory for a massive free particle moving on the
group manifold $AdS_3 \cong SL(2, \mathbb{R} )$ is analysed in
detail. In particular a symplectic structure and two different sets of
canonical coordinates are explicitly found, corresponding to the
Cartan and Iwasawa decomposition of the group. Canonical quantization is
performed in two different ways; by imposing the future-directed constraint
before and after quantization. It is found that this leads to different quantum theories. The Hilbert space of either theory decomposes
into the sum of certain irreducible representations of $sl(2,\mathbb{R}) \oplus
sl(2,\mathbb{R})$; however, depending on how the constraint is imposed
we get different representations. Quantization of the mass occurs,
although a continuum exists in the unconstrained theory corresponding
to particles that can reverse their direction in time.  A quantization in terms
of the ``chiral'' variables of the theory is also carried out giving
the same results. Comparisons are made between QFT in $AdS_3$ and the
quantum mechanics derived, and it is found that one of the quantum
theories is consistent with the Breitenlohner-Freedman bound.}

\begin{document}

\section{Introduction}

Our present understanding of string theory in curved backgrounds is rather
limited. One of the simplest non-trivial backgrounds would appear to
be $AdS_3$; this space is actually the group manifold
$SL(2,\mathbb{R})$, therefore this allows one to re-express bosonic
string theory with a $B$ field as an $SL(2,\mathbb{R})$ WZW model,
which arguably is easier to study. It is important to understand strings in
$AdS_3$ from the point of view of the $AdS/CFT$ correspondence, as a
detailed understanding of the string theory would allow non-trivial
tests of the conjecture and maybe even insights into its proof. Now,
much work has been done on the subject, for example~\cite{mal},~\cite{per},~\cite{bal}, however a
systematic canonical treatment is lacking.
An obvious starting point for all this is to study the simpler case of
a free particle in such a space, since it should correspond to the
$\alpha' \to 0$
limit of the string theory, and of course it is interesting in its
own right. The particle Lagrangian will be cast in a form similar to
the WZW model, and then analyzed in a group theoretic way, following
~\cite{god1},~\cite{fad2}. 

In this paper we first study the classical theory for a massive free particle moving on the group
manifold $AdS_3 \cong SL(2, \mathbb{R} )$. We will make use of the
elegant approach to phase space and Poisson brackets introduced by
Witten~\cite{wit} and Zuckerman~\cite{zuc}. This involves the identification of the phase space with the manifold of all classical
solutions, and defining a symplectic form on this manifold directly
from the Lagrangian. This leads to interesting
(quadratic) Poisson brackets in terms of certain ``natural'' variables. As
expected it is found that the current algebras provide a
representation of $sl(2, \mathbb{R}) \oplus sl(2,\mathbb{R})$ the
Lie algebra of the isometry group of the manifold. Quantization of
this system can be done canonically, leading to the result that
the Hilbert space of the theory furnishes the direct sum of
certain irreducible representations of the quantum current algebra which is still
$sl(2,\mathbb{R}) \oplus sl(2,\mathbb{R})$. The irreducible representations turn out to
be most of both discrete series, but the exact representations seem to
depend on how one imposes the constraint that the particle should be
future directed. The consequences of this are that the particle mass
becomes quantized (as expected since there is a closed time-like
direction). Interestingly $C^{0}_j$ in the exceptional
interval appears in the unconstrained system leading to a continuum of
mass states which enjoy the property of being able to reverse their
direction in time. We must note that quantization of a
particle on $AdS_3$ has already been attempted~\cite{raif}, see also~\cite{fulop}, however
our results differ slightly; if we impose the constraint before
quantization we do indeed reproduce their results, however quantizing
the unconstrained theory we get a quantum correction to the
Casimir which changes the allowed representations, and hence the
masses, even after the constraint is imposed in the quantum theory.

A more interesting quantization, going back to Faddeev et
al.~\cite{fad1},~\cite{fad2} and later Goddard et al.~\cite{god1},~\cite{god2}, is
carried out too, involving quantities which correspond to the ``chiral''
nature of the theory inherited from the bivariance of the metric. This gives the same results, as one of our
quantum theories, and provides a
strategy which will be useful for quantization of the string.

In the final section we compare known results of QFT in $AdS_3$~\cite{satoh},
namely the Wightman function, to the propagator in quantum
mechanics. The propagator can take two different forms, depending on
which expression for the Casimir is chosen. The Casimir with the
quantum correction leads to a simplification in the formulae and thus
appears critical in some sense; further, it is also consistent with the Breitenlohner-Freedman bound in $AdS_3$.

\section{The classical theory}

\subsection{The geometry of $SL(2,\mathbb{R})$}

The Lie algebra $sl(2,\mathbb{R})$ consists of all real two
dimensional traceless matrices. A convenient basis is given by,

\begin{displaymath}
\mathsf{T_1} = \left( \begin{array}{cc}
 0 & 1 \\
 1 & 0
\end{array} \right)
\qquad \mathsf{T_2} = \left( \begin{array}{cc}
 0 & -1 \\
 1 & 0
\end{array} \right)
\qquad \mathsf{T_3} = \left( \begin{array}{cc}
 1 & 0 \\
 0 & -1
\end{array} \right)
\end{displaymath}

which satisfy $\mathsf{T_a} \mathsf{T_b} = \eta_{ab} \mathsf{1} +
\epsilon_{ab}^{\phantom{ab}c} \mathsf{T_c}$, where $(\eta_{ab}) =
\diag(1,-1,1)$ and $\epsilon_{abc}$ is the usual alternating symbol
($\epsilon_{123}=1$), and $\epsilon_{ab}^{\phantom{ab}c} =
\epsilon_{abd} \eta^{dc}$. Note that the structure
constants in this basis are $2\epsilon_{ab}^{\phantom{ab}c}$. The
Killing form of a Lie algebra $\mathfrak{g}$ is defined as a
$(0,2)$ tensor $\kappa$, such that $\kappa (x,y) = \tr[ ad_x ad_y]$
where $x,y \: \in \: \mathfrak{g}$. It is easy to show that
$\kappa (\mathsf{T}_a , \mathsf{T}_b ) = 8\eta_{ab}$, and it is
this metric (actually $\frac{1}{8}\kappa$)  on the Lie algebra
which is used to raise and lower Lie algebra indices.

Now, any group element can be written uniquely as
$\mathsf{g}=e^{u\mathsf{T_2}}e^{\rho
\mathsf{T_1}}e^{v\mathsf{T_2}}$, where $t=u+v, \phi=v-u \: \in \: [0,2\pi)$
and $\rho \geq 0$, this is called the Cartan decomposition. Any (semi-simple) Lie group possesses a natural left and right
invariant metric, $G_{\mu
\nu}=\frac{1}{2}\tr(\mathsf{g}^{-1}\partial_\mu \mathsf{g} \,
\mathsf{g}^{-1}\partial_\nu \mathsf{g})$. A calculation then
gives,

\begin{equation}
\label{metric}
G = -\cosh^2\rho \: dt^2 + d\rho^2 + \sinh^2\rho \: d\phi^2.
\end{equation}

This is the metric for $AdS_3$ with the
cosmological constant
$\Lambda = -1$. To get the metric for the universal cover
$\widetilde{AdS_3}$ one simply takes $t$ to range over $\mathbb{R}$.

\subsection{Massive Particles}

Let $\mathsf{g} : \mathbb{R} \rightarrow SL(2, \mathbb{R} )$ be
the curve corresponding to the particle's worldline. The parameter of
the curve $\mathsf{g}$ will be called the proper time and will be
assumed to take all real values since $AdS_3$ is geodesically complete. The
Lagrangian we will use is,

\begin{equation}
\label{lagrangian}
\mathcal{L} = \frac{1}{2} \tr(\mathsf{g}^{-1}\dot{\mathsf{g}})^2.
\end{equation}

Hence we see, from (\ref{metric}), that the Lagrangian is simply
 $\mathcal{L} = G_{\mu \nu} \dot{g^{\mu}} \dot{g^{\nu}}$, where
$G_{\mu\nu}$ is the metric for $AdS_3$, although it must be
 supplemented with the mass-shell constraint (see Appendix B).

Now, one may wonder why we
have chosen to use (\ref{lagrangian}) as our particle Lagrangian, and
not the more familiar
 $\mathcal{L}=-2m\sqrt{|G_{\mu\nu}\dot{g}^{\mu}\dot{g}^{\nu}|}$, where
$m$ is the mass; this latter Lagrangian is singular (i.e. $ \det
 \big[ \frac{\partial^2 \mathcal{L}}{\partial \dot{g}^{\mu} \partial
 \dot{g}^{\nu}} \big]=0$ and hence one cannot express $\dot{g}^{\mu}$ in
 terms of the canonical momenta), which will lead to a constraint
 (the mass-shell condition) which by itself is not so bad. What
complicates matters is that one cannot gauge fix the reparameterisation
invariance covariantly (see~\cite{hrt} for a good account); breaking spacetime
covariance is undesirable, as after quantization one needs to
show that the theory is still covariant.
 Also it must be noted that the ``square-root
 Lagrangian'' is actually classically equivalent to $\mathcal{L}=
 \frac{1}{e}G_{\mu \nu} \dot{g^{\mu}} \dot{g^{\nu}} - em^2$, where $e$
 is an einbein on $\mathbb{R}$; this Lagrangian can be gauge fixed
covariantly, e.g. $e\approx 1$ \footnote{ The symbol $\approx$ means
weakly equal in the language of Dirac, in other words equal
modulo the constraints.}, and in fact with this constraint we get
that the phase space structure is the same as that of the Lagrangian
(\ref{lagrangian}) with the mass-shell constraint (see Appendix B).

It will be useful to find the Hamiltonian
 formulation for our Lagrangian system. Firstly, the canonical momenta
 are
 $\pi_{\mu} \equiv \frac{\partial \mathcal{L}}{\partial \dot{g}^{\mu}}
 = 2G_{\mu \nu}(g)\dot{g}^{\nu}$. The canonical Hamiltonian is then
 found to be $H=\frac{1}{4} G^{\mu\nu}\pi_{\mu}\pi_{\nu}$. The mass-shell constraint is $\frac{1}{4}G^{\mu\nu}\pi_{\mu}\pi_{\nu}+m^2 \approx 0$. Now at this point one is tempted to take
 the phase space as $T^{*}(AdS_3)$ with coordinates
 $(g^{\mu},\pi_{\nu})$, and proceed as usual by defining the standard
 Poisson bracket, working out the Noether currents and their algebras
 in preparation for the quantum theory. However this route will turn
 out to be plagued with pitfalls at the quantization stage, such as
 operator ordering ambiguities in expressions involving the metric in
 a general coordinate system. Also, solving the equations of motion
 covariantly is not really possible in this formalism. Since we are dealing with a group
 manifold we will employ a more group theoretic approach.

The left invariance of the
metric gives the current $\mathsf{L}= -\dot{\mathsf{g}}
\mathsf{g}^{-1}$ and the right invariance gives $\mathsf{R} =
\mathsf{g}^{-1} \dot{\mathsf{g}}$. The equations of motion are
simply the current conservation laws $ \dot{\mathsf{L}} =
\dot{\mathsf{R}} = 0$. A general solution can be easily obtained in this
language, $\mathsf{g}(\tau)= e^{-\mathsf{L}\tau}\mathsf{g}(0)=\mathsf{g}(0)e^{\mathsf{R}\tau}$.

A more useful form for a general
time-like (it is here that restriction to massive particles occurs) solution can be derived as follows. Firstly note that
isometries map time-like geodesics into time-like geodesics. The action of the
isometries on geodesics is transitive. Hence given one time-like
geodesic, $\mathsf{g}(\tau) = e^{ p\tau \mathsf{T_2}}$ say ($p>0$ for
future-directed),
all others are given by the action of an isometry on it. Therefore
a general time-like geodesic can be written as
$\mathsf{g}(\tau)=\mathsf{u}_0e^{p\tau\mathsf{T}_2}\mathsf{v}_0$, where
$\mathsf{u}_0$ and $\mathsf{v}_0$ are group elements. It is clear that
the map $\mathsf{u}_0 \mapsto \mathsf{u}_0 \mathsf{h}$ and $\mathsf{v}_0
\mapsto \mathsf{h}^{-1} \mathsf{v}_0$, with
$\mathsf{h}=e^{a\mathsf{T}_2}$, leaves $\mathsf{g}(\tau)$ unchanged
(in fact for any such map which leaves $\mathsf{g}(\tau)$ unchanged,
$\mathsf{h}$ must be of this form);
therefore,

\begin{equation}
\mathsf{g}(\tau)= \mathsf{\tilde{u}}
e^{(q+p\tau)\mathsf{T_2}} \mathsf{\tilde{v}}
\label{sol}
\end{equation}

where $\mathsf{\tilde{u}}$ and $\mathsf{\tilde{v}}$  belong to 
$SL(2,\mathbb{R})/T$ and $T \backslash SL(2,\mathbb{R})$
respectively, where $T$ is the Cartan subgroup given by $T = \{ e^{a
\mathsf{T_2}} \}= SO(2)$. Note that using (\ref{sol}) the Hamiltonian
works out to be $H=-p^2 \approx - m^2$.

\subsection{Phase space and Poisson brackets}

We will follow the elegant approach of Zuckerman~\cite{zuc} and Witten~\cite{wit}, in order to get a
covariant derivation of the Poisson brackets. This involves defining
the phase space $\mathcal{S}$, as the manifold of classical solutions
of the Euler-Lagrange equations of the Lagrangian. Then in order to
define Poisson brackets, as is well known, one needs to find a
symplectic form on $\mathcal{S}$; recall this is simply a closed
non-degenerate two form. The idea is that one can avoid moving into
the Hamiltonian formalism by defining a symplectic form directly from
the Lagrangian $\mathcal{L}$; we sketch how this works in the general case of a
one-dimensional field theory. Consider a theory with fields $\phi : \mathbb{R}
\rightarrow M$ and an action $S(\phi)= \int_{\mathbb{R}}
\mathcal{L}(\phi)$. Let $\mathcal{F}$ be the space of fields $\phi$ and
$\mathcal{S}$ the submanifold of solutions to the variational problem
$\delta S(\phi)=0$, where $\delta$ is the exterior derivative on
$\mathcal{F}$. Now, $\mathcal{L}$ is a (0,1)-form on $\mathcal{F}\times\mathbb{R}$; it can be shown that $\delta \mathcal{L} = E +d\theta$ where
$d$ is the exterior derivative on $\mathbb{R}$ and the decomposition is unique. Then one defines the two form (on
$\mathcal{S}$) as $\omega=\delta \theta (\tau)$ which is the symplectic
structure desired; it is clear, since $d$ and $\delta$ anticommute on
$\mathcal{F} \times \mathbb{R}
$, and that $E$ vanishes on $\mathcal{S}\times \mathbb{R}$ (this corresponds to the
Euler-Lagrange equations as $\delta S(\phi)= \int_{\mathbb{R}}E(\phi,\delta\phi)$) that $d\omega = 0$ showing that no
special choice of $\tau$ has been made.

Applying this procedure to (\ref{lagrangian}) gives $\theta(\tau) =-
\tr[\delta \mathsf{g} \; \partial_{\tau}(\mathsf{g}^{-1})]$. Therefore a symplectic form for $\mathcal{S}$ is given by,

\begin{equation}
\label{symplectic}
\omega = -\frac{1}{2} \tr \big( \delta \mathsf{g} \wedge \delta (
\partial_{\tau}\mathsf{g}^{-1}) \big)
\end{equation}

where the factor of two has been introduced for convenience.

Poisson brackets for the
theory are defined in the usual way,

\begin{displaymath}
\{ f,g \} = \omega^{ab} \partial_a f \, \partial_b g, \qquad
\textrm{where} \quad f,g \: \in \: C^\infty (\mathcal{S}) \qquad
\textrm{and} \quad \omega_{ac} \omega^{cb} = \delta^{a}_b.
\end{displaymath}

Now, one can choose a parameterisation of the
cosets such that $\mathsf{\tilde{u}}= e^{A\mathsf{T_2}}e^{B\mathsf{T_1}}$
and $\mathsf{\tilde{v}}=e^{\bar{B} \mathsf{T_1}}e^{\bar{A} \mathsf{T_2}}$.
Note that choosing a different parameterisation simply redefines
$q$ appearing in (\ref{sol}). Using this parameterisation, and substituting into (\ref{symplectic}) we
get,

\begin{equation}
\omega = \delta \lambda, \quad \textrm{where} \quad \lambda = p\delta
q + M\delta A + \bar{M}\delta \bar{A}
\end{equation}

and $ M=p\cosh2B$ and $\bar{M}=p\cosh2\bar{B}$. Hence we have found
canonical coordinates on the phase space with  only the
following non-zero Poisson brackets,

\begin{equation}
\{ q,p \} = \{ A,M \} = \{ \bar{A}, \bar{M} \} = 1.
\end{equation}

It is also interesting to work out Poisson brackets of other
variables. Define $\mathsf{u} =
\tilde{\mathsf{u}}e^{q\mathsf{T_2}}$ and $\mathsf{v} =
e^{q\mathsf{T_2}} \tilde{\mathsf{v}}$, and use the following
convenient notation \footnote{Define $(\mathsf{a \otimes
b})_{ik,jl} = (\mathsf{a})_{ij} \, (\mathsf{b})_{kl}$, and
$\mathsf{(u)}_{ik,pq} \mathsf{(v)}_{pq,jl} = \mathsf{(u \,
v)}_{ik,jl}$. Then we get nice laws such as $\mathsf{(a \otimes
b)(c \otimes d)} = \mathsf{a c \otimes b d}.$} for $\mathsf{a} \:
\in \: \Mat(2,\mathbb{R})$, $\mathsf{a_1}=\mathsf{a} \otimes
\mathsf{1}$ and $\mathsf{a_2} = \mathsf{1} \otimes \mathsf{a}$.
Then it is found that,

\begin{eqnarray}
\label{uvalg}
\{ \mathsf{\tilde{u}_1} , \mathsf{\tilde{v}_2} \} = \{
\mathsf{\tilde{u}}, p \} =    \{ \mathsf{\tilde{v}}, p \} = 0 \\
\{ \mathsf{u_1}, \mathsf{u_2} \} = \mathsf{u_1 u_2 r_{12}} \\   \{
\mathsf{v_1}, \mathsf{v_2} \} = - \mathsf{r_{12} v_1 v_2} \\
\mathsf{r_{12}} = \frac{1}{2p} ( \mathsf{T_1 \otimes T_3 - T_3
\otimes T_1} ). \label{rmatrix}
\end{eqnarray}

Now, the currents are easily found to be,
\begin{eqnarray}
\mathsf{L} = - \mathsf{\tilde{u}}p\mathsf{T_2 \tilde{u}^{-1}} = -
\mathsf{u}p\mathsf{T_2 u^{-1}} \\ \mathsf{R} =
\mathsf{\tilde{v}^{-1}} p\mathsf{T_2 \tilde{v}} = \mathsf{v^{-1}}
p\mathsf{T_2 v}
\end{eqnarray}

which then give the following brackets

\begin{eqnarray}
\label{lurv}
\{ \mathsf{L_1} , \mathsf{u_2} \} = - \mathsf{\mathcal{C}_{12}
u_2}, \qquad \{ \mathsf{R_1} , \mathsf{v_2} \} =  \mathsf{v_2
\mathcal{C}_{12}}, 
\end{eqnarray}

where $\mathsf{\mathcal{C}}_{12} = \mathsf{T}^a
\otimes \mathsf{T}_a$ is the tensor Casimir. This gives us the current algebras,

\begin{equation}
\{\mathsf{L_1} , \mathsf{L_2} \} = [\mathsf{L_2} ,
\mathsf{\mathcal{C}_{12}}], \qquad \{\mathsf{R_1} , \mathsf{R_2}
\} = [\mathsf{R_2} , \mathsf{\mathcal{C}_{12}}].
\end{equation}

Since $\mathsf{L}$ and $\mathsf{R}$ belong to the Lie algebra
$sl(2,\mathbb{R})$, then we can write the current algebras in
terms of their components in the basis $\{ \mathsf{T}_a \}$. We
get,

\begin{equation}
\{ L_a ,L_b \} = 2\epsilon_{ab}^{\phantom{ab}c} L_c, \qquad
\{R_a,R_b \} = 2\epsilon_{ab}^{\phantom{ab}c} R_c
\end{equation}

which are simply two copies of $sl(2,\mathbb{R})$. It should be noted
that the current algebras can be deduced from equations (\ref{uvalg})-(\ref{rmatrix}) alone
without the need to resort to canonical coordinates. A possible
sticking point in the derivation of (\ref{lurv}) appears to be when one gets
to,

\begin{equation}
\label{lucalc}
\{\mathsf{L}_1 , \mathsf{u}_2 \} =
-\mathsf{u_1u_2\mathcal{C}_{12}u_1^{-1}}.
\end{equation}
 It seems that the only way to know how $\mathsf{\mathcal{C}_{12}}$
 moves past the $\mathsf{u}$'s is to use an explicit parameterisation
 (and hence canonical coordinates). In fact we can avoid working in
 canonical coordinates as follows. Consider
 $\mathbb{P}= \frac{1}{2}( \mathsf{I \otimes I} +
 \mathsf{\mathcal{C}}_{12})$. It is convenient to use the isomorphism
 $\Mat(n,\mathbb{R}) \otimes \Mat(n,\mathbb{R}) \cong
 \Mat(n^2,\mathbb{R})$ defined by,

\begin{displaymath}
\left( \begin{array}{cc} a & b \\ c & d \end{array} \right)
\otimes  \left( \begin{array}{cc} \alpha & \beta  \\ \gamma & \delta
\end{array} \right)  \mapsto \left( \begin{array}{cccc} a\alpha & a\beta &
b\alpha & b\beta \\ a\gamma & a\delta & b\gamma & b\delta \\ c\alpha & c\beta &
d\alpha & d\beta \\ c\gamma & c\delta & d\gamma & d\delta \end{array}
\right)
\end{displaymath}

for $n=2$. Then it is easy to show that,

\begin{eqnarray}
\mathsf{\mathcal{C}}_{12} = \left( \begin{array}{cccc} 1 & 0 & 0 & 0 \\ 0 &
-1 & 2 & 0 \\ 0 & 2 & -1 & 0 \\ 0 & 0 & 0 & 1 \end{array} \right) \\
\mathbb{P} = \left( \begin{array}{cccc} 1 & 0 & 0 & 0 \\ 0 &
0 & 1 & 0 \\ 0 & 1 & 0 & 0 \\ 0 & 0 & 0 & 1 \end{array} \right)
\end{eqnarray}

Note that  $\mathbb{P}^2 = \mathsf{I \otimes I}$. If we consider $\mathsf{a}
\in \Mat(2,\mathcal{A})$ where $\mathcal{A}$ is a not necessarily commutative
algebra, then it is a straightforward computation to check that,

\begin{equation}
\label{perm}
\mathbb{P}\mathsf{a}_1\mathsf{a}_2 =
\mathsf{a}_2\mathsf{a}_1\mathbb{P} \label{P}
\end{equation}

which reduces to $\mathsf{\mathcal{C}}_{12} \mathsf{a}_1\mathsf{a}_2 =
\mathsf{a}_1\mathsf{a}_2 \mathsf{\mathcal{C}}_{12}$ in the case where
$\mathcal{A}$ is commutative. Applying this to (\ref{lucalc}) (where $\mathcal{A}
= \mathbb{R}$) we get the desired result. Equation (\ref{perm}) will be needed
in the quantum theory.

\subsection{Global coordinate systems and constraints}

In the previous section we parameterised the cosets using the Cartan
decomposition, leading to a global set of coordinates
$(q,p,A,M,\bar{A},\bar{M})$ for the phase space. Actually, this is not
quite true since $A$ and $\bar{A}$ are identified modulo $2\pi$, and
hence as functions $S^1 \to \mathbb{R}$ are not continuous and hence
not smooth. This corresponds to the well known fact that one cannot cover a
circle with a single chart. Thus, strictly, the formulas involving
these coordinates at best are valid in some open region of the
circles, for example everywhere except one point, such as $2\pi$. A
set of Poisson brackets which do hold globally are $\{ \sin A , M \} =
\cos A$ and $\{ \cos A, M \} = -\sin A$, and these serve as a global
replacement of $\{A,M\}=1$, which as we have explained can only be
satisfied at best everywhere except at a point. Of course all the
formulas derived actually involve only $\sin A$ or $\cos A$ and hence
are all valid globally. See Isham~\cite{isham} for a good account of
such subtleties.

Now it is clear that the future-directed constraint $p>0$ will need to
be imposed. This can be either done before or after quantization. In
the coordinates developed so far this is best done after. This is because
restricting the phase space classically to $p>0$, which implies $M>0$
and $\bar{M}>0$, means that we will have to quantize canonical
variables of the form $S^1 \times \mathbb{R}^+$ which is problematic. Instead we will derive a set of
coordinates which after the restriction $p>0$ still allow an easy
quantization.

To do this, we need the Iwasawa decomposition of
$SL(2,\mathbb{R})$. This tells us that any element of the group
manifold can be written as
$\mathsf{g}=e^{A\mathsf{N}_+}e^{B\mathsf{T}_3}e^{C\mathsf{T}_2}$,
where $\mathsf{N}_+=\left( \begin{array}{cc} 0 & 1 \\ 0 & 0 \end{array} \right) $, and thus the
coset $G / H$ can be parameterised by $\mathsf{\tilde{u}}=e^{A\mathsf{N}_+}e^{B\mathsf{T}_3}$. If
one proceeds to calculate the symplectic form, we get the same
expression as before except that $M=-\frac{1}{2}pe^{-2B}$ and
similarly for $\bar{M}$. However, now the variables $A,\bar{A}$ are
not periodic but take all real values. This means that if one imposes
the constraint classically the phase space restricts to three
canonical pairs of the form $\mathbb{R}^+ \times \mathbb{R}$ or
$\mathbb{R}^- \times \mathbb{R}$ which are both straightforward to deal
with since both these are symplectomorphic to the standard $\mathbb{R}
\times \mathbb{R}$ phase space.
More explicitly, one finds that the current $\mathsf{L}=-\mathsf{\dot{g}g^{-1}}$ has the following components in these Iwasawa coordinates,

\begin{eqnarray}
L^3=2AM \\
L^+=L^1-L^2= -\left( \frac{p^2}{2M}+2A^2M \right) \\
L^-=L^1+L^2= 2M
\end{eqnarray}

and of course it is easy to verify from the canonical Poisson brackets
that they satisfy the $sl(2,\mathbb{R})$ algebra. Analogous
expressions hold for the right current $\mathsf{R}$.

\section{The quantum theory}

\subsection{Canonical quantization}

Canonical quantization has a long history and many problems. Schematically, given a phase
space $\mathcal{S}$ and canonical coordinates $(q_i ,p_i)$, then the
quantum theory is constructed via a correspondence map
$\hat{\phantom{a}} : C^{\infty}(\mathcal{S}) \rightarrow \End \,
\mathcal{H}$, such that the quantum observables are Hermitian with
respect to the inner product on $\mathcal{H}$. The canonical
coordinates will satisfy $[\hat{q_i},\hat{p_j}] = i\hbar
\delta_{ij}$, and
\[\lim_{\hbar \rightarrow 0} \frac{[\hat{f} , \hat{g}]}{i\hbar} =
\widehat{\{f,g \}}\] for more general observables. Now this only works
if the canonical coordinates take all real values, and is not
particularly useful unless they are global coordinates for the phase
space. It is also clear
that ordering ambiguities arise upon quantization, and in general
these have to be resolved case by case.

Before moving on we recall the standard, straightforward example of
when the phase space is $\mathbb{R}^n \times \mathbb{R}^n$. Then we take global coordinates
$(q_i, p^i)$ and of course the natural symplectic form
$\omega=\sum_i dp^i \wedge dq_i$ gives the Poisson brackets
$\{q_i,p^j\}=\delta_{i}^{j}$. To quantize the system one takes unitary
representations of this algebra, called the Heisenberg algebra, and it
is well known that there exists a unique irreducible representation of
this realised on $L^2(\mathbb{R}^n)$ (in fact the more precise
statement concerns unitary representations of the Heisenberg group
since this will involve strictly bounded operators, see~\cite{isham}).

\subsubsection{Cartan coordinates}

If we carry this procedure out for the system studied in the
previous section then we simply get the following quantum
conditions \footnote{We make use here of the correct quantization of
the angular variables, as discussed in the previous section and in
more detail in ~\cite{isham}.},

\begin{equation}
[q,p]=i\hbar, \qquad [e^{iA},M]=-\hbar e^{iA}, \qquad
[e^{i\bar{A}},\bar{M}]=-\hbar e^{i\bar{A}}.
\end{equation}

The Hamiltonian will be $H=-p^2$, and a
general observable, $O$, will evolve according to Heisenberg's equation
of motion $i\hbar \dot{O}= [O,H]$, which
is equivalent to $O(\tau)= e^{iH \tau / \hbar}O(0) e^{-iH \tau /
\hbar}$. Of course we are now interested in the quantum versions of
the current algebras; tentatively  we take the left current to be $\mathsf{L}=-\tilde{\mathsf{u}}p\mathsf{T}_2
\tilde{\mathsf{u}}^{-1}-i\alpha_{L}\hbar\mathsf{I}$ where $\alpha_{L}$ is some
unknown constant. Now, if we compute the components of the current
we find that indeed there are potential ordering problems since we end
up with functions of $A$ times functions of $M$, however they happen
to be all Hermitian combinations of the canonical variables, and so we
conclude that we have taken an acceptable definition for the
current. Note that if we choose $\alpha_{L}=1$ we get $\tr\mathsf{L}=0$
which is a desirable property (but not essential). Anyway, we get

\begin{equation}
L_2 = M, \qquad L_{\pm} = -e^{\pm 2iA}\sqrt{(M \pm
\hbar)^2 -p^2}
\end{equation}

where $L_{\pm} = L_3 \pm i L_1$. It is easy to the check that $<L_1,L_2,L_3>$ provide a unitary
representation of the
algebra of $sl(2,\mathbb{R})$,

\begin{equation}
[L_2,L_{\pm}]=\pm 2\hbar L_{\pm}, \qquad [L_+,L_-]=-4\hbar L_2.
\end{equation}

The Casimir
$Q_L=\eta^{ab}L_aL_b$ can be also calculated to give,

\begin{equation}
\label{casimir}
Q_L = -p^2 + \hbar^2.
\end{equation}

A similar set of calculations for the right algebra
gives,

\begin{equation}
Q_R = -p^2 + \hbar^2.
\end{equation}

Therefore we deduce that $Q_L=Q_R$ which then tells us that the
Hilbert space of states for the quantum theory decomposes as
follows,

\begin{equation}
\mathcal{H}= \bigoplus_{\varrho} \mathcal{V^{L}}_{\varrho} \otimes
\mathcal{V^{R}}_{\varrho}
\end{equation}

where $\varrho$ is the eigenvalue of the Casimir which labels the
irreducible representations, and $\mathcal{V^{L,R}}_{\varrho}$ is the carrier space of an
irreducible representation. Explicitly (for the left algebra) we have,
\begin{eqnarray}
Q_L|\varrho ,m\rangle = 4\hbar^2 \varrho |\varrho,m\rangle \\
L_2 |\varrho,m\rangle = 2\hbar m|\varrho,m\rangle \\
L_{\pm}|\varrho,m\rangle = 2\hbar \sqrt{\varrho+ m(m \pm 1)} |\varrho,m \pm 1\rangle
\end{eqnarray}
where $\{ |\varrho,m\rangle  \}$ is a basis
for $\mathcal{V^L}_{\varrho}$, and $m$ is either integer or half
integer depending on the representation. However not all irreducible representations of $sl(2,\mathbb{R})$ are allowed.
In fact the only ones allowed are $D^{+}_j$, $D^{-}_j$ 
(where
$\varrho=-j(j+1), \: j \in \{ -1, -3/2,... \}  $) and $C^{0}_{\varrho}$ for
$0< \varrho< \frac{1}{4}$. This follows from the fact that $Q_L < \hbar^2$,
and hence $\varrho < \frac{1}{4}$.

We deduce that the allowed values for the mass (in units of $\hbar$)
\footnote{The geodesic $\mathsf{g}(\tau)=
e^{(q+p\tau)\mathsf{T}_2}$ in the coordinates
$x^{\mu}=(t,\rho,\phi)$ is $g^{\mu}(\tau)=(q+p\tau,0,\phi_0)$.
Hence $\dot{g}^{\mu}= (p,0,0)$, and thus $m^2=\mu^2\hbar^2\equiv -G_{\mu\nu}
\dot{g}^{\mu} \dot{g}^{\nu} = p^2$. Of course the mass $\mu \hbar$ is then defined
as the positive square root. This is all classical; Appendix B tells
us that the same holds quantum mechanically i.e. $p^2=m^2$ on physical states.} are,

\begin{eqnarray}
\mu =  |2j+1|, \qquad j \in \{ -1, -3/2,... \} \\
\mu = \sqrt{1-4\varrho}, \qquad 0< \varrho <\frac{1}{4},
\end{eqnarray}

that is $\mu$ is either a positive integer or in the interval $(0,1)$.

Quantization of the mass is not unexpected since $AdS_3$ has
closed time-like direction, that is topologically it is $S^1
\times \mathbb{R}^2$, where the $S^1$ refers to time. The continuum
 $0<\mu<1$ is more interesting.

An explicit basis for $\mathcal{H}$ can be constructed from a state
$|0\rangle$ satisfying $p|0\rangle= M|0\rangle = \bar{M}|0\rangle =
0$. Namely $\{ |\mu,m,\bar{m}\rangle \}$, where

\begin{equation}
|\mu,m,\bar{m}\rangle = e^{i\mu q} e^{2imA} e^{2i\bar{m}\bar{A}} |0\rangle
\end{equation}

will span the Hilbert space for appropriate values of $\mu,m,\bar{m}$. It
is in fact an orthogonal basis and normalised if $\langle0|0\rangle =1$.

We have not yet imposed the constraint $p>0$ corresponding to the
future directed condition of the geodesics. Now we need to find a way
to impose this condition quantum mechanically. This turns out to be
actually fairly easy. We merely note that classically the constraint
told us that $M>0$. Hence we see that quantum mechanically a sensible
constraint is $L_2>0$; note the corresponding statement $R_2<0$ also
exists. Thus we need to pick a subspace of our Hilbert space for which
these conditions hold. This is easy to do and gives,

\begin{equation}
\mathcal{H}_{p>0} = \bigoplus_{j\leq -1} D^+_j \otimes D^-_j.
\end{equation}

Note in particular that the mysterious continuum has
disappeared. This continuum actually corresponds to states which can
flip their direction in time and of course have no classical
counterpart; naturally they are excluded in the quantum theory after
imposing the future directed constraint.

\subsubsection{Iwasawa coordinates}

In these coordinates we have the interesting option of imposing the
constraint $p>0$ before or after quantization. Interestingly this leads
to slightly different quantum theories.

First we will mimic the previous section and quantize the
unconstrained system and then impose the constraint. Of course we hope
to reproduce the same results. The quantum conditions are,

\begin{equation}
[q,p]=i\hbar, \qquad [A,M]=i\hbar, \qquad [\bar{A},\bar{M} ] =i\hbar,
\end{equation}

and note there are no subtleties due to periodic variables
here. Resolving some of the ordering ambiguities in the currents
(imposing that the components be self-adjoint suffices) leads to our
left current being,

\begin{eqnarray}
L^3=AM+MA \\
L^+=L^1-L^2= -\left( \frac{p^2}{2M}+ 2AMA \right) \\
L^-=2M
\end{eqnarray}
 and these satisfy the same $sl(2,\mathbb{R} )$ algebra as in the
 previous section, as they should! The Casimir can be computed, and
 reassuringly we get $Q_L=-p^2 +\hbar^2$, which is the same as we got
 by quantization of the system in the very different Cartan
 coordinates. Note this wasn't a priori guaranteed, since it isn't
 necessarily obvious that quantizing a system in completely different
 canonical coordinates is compatible. Of course, from here on we'll
 get the same quantum theory as in the previous section before and
 after the future-directed constraint is imposed.

Now, as we've mentioned in these coordinates we can actually quantize
the constrained classical system. To do this, we briefly explain how
one would quantize the classical system with phase space $\mathbb{R}
\times \mathbb{R}^+$, see~\cite{isham}. Let $(x,\pi)$ be global coordinates. There is a
nice way of mapping this to the standard $\mathbb{R}
\times \mathbb{R}$ phase space; simply define the
diffeomorphism $(x,\pi)\to (x\pi,\log \pi) \equiv
(\tilde{x},\tilde{\pi})$. It then follows that $\omega= d\pi \wedge dx
= d\tilde{\pi} \wedge d\tilde{x}$ showing that the phase spaces are
symplectomorphic as previously asserted. Therefore one can choose
$(\tilde{x},\tilde{\pi})$ as canonical coordinates and simply quantize
as usual via a Heisenberg algebra,
$[\tilde{x},\tilde{\pi}]=i\hbar$. Note since $\pi=e^{\tilde{\pi}}$ it
follows that $[\tilde{x},\pi]=i\hbar\pi$, which can be taken as our
fundamental quantum condition for quantum mechanics on
$\mathbb{R}\times \mathbb{R}^+$. Now, lets apply this to our
system. Expressing the classical current $\mathsf{L}$ in components in the
coordinates $(\tilde{q}=qp,p,\tilde{A},M=-e^{\tilde{M}})$ gives,

\begin{eqnarray}
L^3=2\tilde{A} \\
L^+=L^1-L^2= -\left( \frac{p^2}{2M}+\frac{2\tilde{A}^2}{M} \right) \\
L^-=L^1+L^2= 2M.
\end{eqnarray}
 
Quantization then involves imposing the conditions,

\begin{equation}
[\tilde{q},p]=i\hbar p, \qquad [\tilde{A},M]=i\hbar M, \qquad
[\tilde{\bar{A}},\bar{M}]=i\hbar \bar{M},
\end{equation}

and quantization of the currents requires simply resolving the order
ambiguity in $L^+$ which we do by writing
$L^+=-(\frac{p^2}{2M}+2\tilde{A}M^{-1}\tilde{A})$. Then, once again it
is an easy exercise showing that we have the $sl(2,\mathbb{R})$
algebra satisfied using the quantum conditions. Interestingly
computing the Casimir gives $Q_L=-p^2$ in contrast to the previous
quantizations. Now, the Hilbert space decomposes into irreducible representations of
$sl(2,\mathbb{R}) \oplus sl(2,\mathbb{R})$, and it is easy to see that
$L_2>0$ as an operator (since $M<0$). Thus since $Q<0$ (the right
algebra will have the same Casimir) we see that the Hilbert space
decomposes as,

\begin{equation}
\mathcal{H}_{p>0} = \bigoplus_{j\leq -\frac{3}{2}} D^+_j \otimes D^-_j.
\end{equation}

Note this differs from our other quantum theory by the absence of
$D^+_{-1} \otimes D^-_{-1}$ in the Hilbert space; of course we get a
different mass spectrum, namely $\mu=\sqrt{j(j+1)}$, which agrees with~\cite{raif}.



\subsection{``Chiral'' quantization}

In this section we aim to show how a subtly different kind of
quantization can be done, involving quantities which relate more
directly to the symmetries of the system. The proposed
quantization~\cite{fad1},~\cite{fad2},~\cite{god1}, consists of quantizing $q,p$ to Hermitian operators
as before, and quantizing $\mathsf{u}$, $\mathsf{v}$ to operator
matrices which are unitary in the sense $\mathsf{u}_{ab}
\mathsf{u}^{\dagger}_{cb}=\delta_{ac}$ \footnote{Note that
unitarity of $\mathsf{u}$ and $\mathsf{v}$ is consistent with
canonical quantization if, for example, the choices $B^{\dagger}=-B$ and
$\bar{B}^{\dagger}=-\bar{B}$ are made for the Cartan coordinates. This is allowed since the
canonical variables $M$ and $\bar{M}$ are even functions of $B$ and
$\bar{B}$ respectively, and hence their Hermiticity is unaffected by
such a choice.}. The quantum conditions are

\begin{eqnarray}
\label{qc1}
[q,p]=i\hbar, \qquad [\mathsf{\tilde{u}}_1, \mathsf{\tilde{v}}_2 ]=0,
\qquad [\tilde{\mathsf{u}},p]=[\tilde{\mathsf{v}},p]=0 \\ \label{qc2}
\mathsf{u_1u_2} = \mathsf{u_2 u_1 B_{12}}, \qquad \mathsf{v_1v_2 =
B_{12}^{-1} v_2 v_1}
\end{eqnarray}

where $\mathsf{\tilde{u}}=\mathsf{u} e^{-q\mathsf{T}_2}$ and
$\mathsf{\tilde{v}} = e^{-q\mathsf{T}_2} \mathsf{v}$. Quantum
consistency requires that

\begin{eqnarray}
\mathsf{B} = \mathsf{I \otimes I} + i\hbar \mathsf{r}_{12} +
O(\hbar^2), \qquad \textrm{Classical limit} \\ \label{ua}
\mathsf{B}^{-1} = \mathsf{B}^{\dagger} =
\mathbb{P}\mathsf{B}\mathbb{P}, \qquad \textrm{Unitarity and
antisymmetry  }  \\ \label{jac}
\mathsf{B}_{23}(\tilde{p}_1)\mathsf{B}_{13}(p)\mathsf{B}_{12}(\tilde{p}_3)
= \mathsf{B}_{12}(p) \mathsf{B}_{13}(\tilde{p}_2) \mathsf{B}_{23}(p),
\qquad \textrm{Associativity} \\ \label{loc}
\big[\mathsf{B}, e^{q(\mathsf{(T_2)_1 +(T_2)_2})}\big] = 0 \qquad
\textrm{Locality}
\end{eqnarray}

where $\tilde{p}= p - i\hbar \mathsf{T}_2$ (note that $p\mathsf{u} =
\mathsf{u}\tilde{p}$). Equations (\ref{ua}) follow from $\mathsf{u}$ being
unitary and equations (\ref{qc2}). Equation (\ref{jac}) comes from associativity
$\mathsf{u_1(u_2u_3)=(u_1u_2)u_3}$, where $\mathsf{u_1= u\otimes
1\otimes 1}$ and similarly for the others. Equation (\ref{loc})
deserves some attention; it is equivalent to
$[\mathsf{g}_1(\tau),\mathsf{g}_2(\tau) ] =0$ hence ensuring
locality in the quantum theory. We will show this at $\tau=0$, the
general result follows from Heisenberg's equations of motion. First
note that $\mathsf{g}(0)= \mathsf{\tilde{u}} \mathsf{Q}
\mathsf{\tilde{v}} = \mathsf{uQ^{-1}v}=
\mathsf{u\tilde{v}}=\mathsf{\tilde{u}v}$, where $\mathsf{Q}=e^{q\mathsf{T}_2}$. Now, the argument runs as follows:

\begin{eqnarray}
\mathsf{g_1(0)g_2(0)= u_1 \tilde{v}_1 \tilde{u}_2 v_2 = u_1
\tilde{u}_2 \tilde{v}_1 v_2= u_1 u_2 Q_2^{-1} Q_1^{-1} v_1 v_2} \nonumber \\ \nonumber
 \mathsf{= u_2 u_1 B_{12} Q_2^{-1} Q_1^{-1} B_{12}^{-1} v_2 v_1 = u_2 u_1
Q_2^{-1} Q_1^{-1} v_2 v_1} \\ \nonumber \mathsf{ = u_2 \tilde{u}_1 \tilde{v}_2 v_1 = u_2
\tilde{v}_2 \tilde{u}_1 v_1 =g_2(0) g_1(0)} 
\end{eqnarray}

where (\ref{loc}) was used in the fifth equality and the quantum
conditions were used in the other steps. It is easy to see, from the algebra
above, that
requiring $[\mathsf{g}_1(0), \mathsf{g}_2(0)] = 0$ implies
(\ref{loc}), hence completing the equivalence.

Now, an explicit formula for $\mathsf{B}$ the braiding matrix is wanted. If we
start from the ansatz,

\begin{equation}
\mathsf{B}(p)=\exp \left( -\sum_{\alpha \in \{\pm1\} } \theta_{\alpha}(p)
\mathsf{E}_{\alpha} \otimes \mathsf{E}_{-\alpha} \right)
\end{equation}

where $\mathsf{E}_{\pm}= \frac{1}{2}(\mathsf{T}_3 \pm i\mathsf{T}_1)$, and then
impose the conditions above we get $\theta_{-\alpha} =
-\theta_{\alpha}$ (from (\ref{ua}),(\ref{loc}) independently), and that $\theta_{\alpha}$ can be any analytic
function of $p$ or $p^{-1}$ (from (\ref{jac})). A computation gives,

\begin{equation}
\mathsf{B}(p)= \mathsf{I \otimes I} - \sin^2(\theta/2)( \mathsf{1
\otimes 1} + \mathsf{T}_2 \otimes \mathsf{T}_2) -
\sum_{\alpha}\sin\theta_{\alpha} \mathsf{E}_{\alpha} \otimes
\mathsf{E}_{-\alpha}
\end{equation}

where $\theta \equiv  \theta_{+1}$. Imposing the classical limit
implies,

\begin{equation}
\sin\theta_{\alpha} = \frac{\hbar}{p\alpha} + O(\hbar^2).
\end{equation}

If we make the choice $\sin\theta_{\alpha} = \frac{\hbar}{\alpha p}$
then we get,

\begin{equation}
\mathsf{B}(p)= \mathsf{I \otimes I} + i\hbar \,\mathsf{r}(p) -
\frac{1}{2}\bigg( 1- \sqrt{1-\frac{\hbar^2}{p^2}}\bigg)(\mathsf{I \otimes I} +
\mathsf{T}_2 \otimes \mathsf{T}_2). \label{brad}
\end{equation}

Now we come to the problems of ordering of the quantum variables.
In particular how do we define the currents $\mathsf{L,R}$ in the
quantum theory. Motivated by the canonical quantization we define the
quantum currents to be,

\begin{eqnarray}
\mathsf{L}= - \tilde{\mathsf{u}} p\mathsf{T}_2 \tilde{\mathsf{u}}^{-1}
- i\alpha_{L}\hbar\mathsf{I} \\
\mathsf{R}=  \tilde{\mathsf{v}}^{-1} p\mathsf{T}_2 \tilde{\mathsf{v}}
+ i\alpha_{R}\hbar\mathsf{I}
\end{eqnarray}

where $\alpha_{L,R}$ are unknown constants. This now allows one to
calculate the following brackets,

\begin{equation}
[\mathsf{L}_1 , \mathsf{u}_2 ] = -i\hbar \mathsf{\mathcal{C}}_{12}
\mathsf{u}_2, \qquad [\mathsf{R}_1 , \mathsf{v}_2 ] = i\hbar
\mathsf{v}_2 \mathsf{\mathcal{C}}_{12} \label{lu}
\end{equation}

\begin{equation}
[\mathsf{L}_1,
\mathsf{L}_2]=i\hbar[\mathsf{L}_2,\mathsf{\mathcal{C}}_{12}], \qquad [\mathsf{R}_1,\mathsf{R}_2]=i\hbar[\mathsf{R}_2,\mathsf{\mathcal{C}}_{12}]
\end{equation}

which reduce to the algebras of $sl(2,\mathbb{R})$ when one expands
$\mathsf{L}= L^a \mathsf{T}_a + k\mathsf{I}$, and similarly for $\mathsf{R}$. What is remarkable is
that when one computes the Casimirs $Q_L = \eta^{ab}L_aL_b$ and $Q_R=\eta^{ab}R_aR_b$, one finds
that they are independent of $\alpha_L$ and $\alpha_R$, and we get,

\begin{equation}
Q_L = -p^2 + \hbar^2, \qquad Q_R = -p^2 +\hbar^2
\end{equation}

However to do this the result
$\tr(\tilde{\mathsf{u}}p\mathsf{T}_2\tilde{\mathsf{u}}^{-1}) =
-2i\hbar$ is needed, which appears only to be calculable using
canonical coordinates (we used the Cartan coordinates to do this). Note that to show equations (\ref{lu}) the
explicit form of the braiding matrix (\ref{brad}) is required.
Straightforward manipulations lead to,

\begin{equation}
[\mathsf{L}_1 , \mathsf{u}_2 ] = \mathsf{u_1u_2} [\mathsf{B}_{12}^{-1}
- \mathsf{I \otimes I}, p(\mathsf{T}_2)_1 ] \mathsf{u}_{1}^{-1}
+i\hbar\mathsf{u_1u_2} \mathsf{(T_2 \otimes T_2)}
\mathsf{B}^{-1}_{12}\mathsf{u}^{-1}_1 \label{lumess}
\end{equation}

and from equation (\ref{brad}) it is easily verified that,

\begin{eqnarray}
[\mathsf{B}_{12}^{-1} - \mathsf{I \otimes I}, p(\mathsf{T}_2)_1 ] =
-i\hbar(\mathsf{T_1 \otimes T_1 + T_3 \otimes T_3}) \\
(\mathsf{T_2 \otimes T_2})(\mathsf{B}_{12}^{-1} - \mathsf{1 \otimes
1}) = \mathsf{B}_{12}^{-1} - \mathsf{I \otimes I}
\end{eqnarray}

which when substituted into (\ref{lumess}) give,

\begin{equation}
[\mathsf{L}_1 , \mathsf{u}_2 ] =
-2i\hbar\mathsf{u_1u_2}\mathbb{P}\mathsf{u}^{-1}_1 +i\hbar
\mathsf{u}_2.
\end{equation}

Now using equation (\ref{P}) where $\mathcal{A} = \End\mathcal{H}$ we get
the first of equation (\ref{lu}) as required, by the sole use of the quantum
conditions (\ref{qc1}), (\ref{qc2}) and (\ref{brad}).

\section{Wavefunctions and comparison to QFT}

It is interesting to see how the quantum mechanics derived
compares to QFT in $AdS_3$ which, of course, has been well
studied, for example in~\cite{satoh}. More precisely, we will compare the
propagator in quantum mechanics to the Wightman function in the
QFT.

 In the quantum mechanics we look for position eigenstates $|g
\rangle$ satisfying $\hat{g} |g \rangle = g|g\rangle$, where we have
put a hat on the operator for clarity. Note that it is a nontrivial
fact that we can simultaneously diagonalise the matrix elements of
$\hat{g}$, but in fact one can, since they all commute amongst each
other. This is the locality condition addressed in the ``Chiral
quantization'' section. The quantity we are interested in then is the amplitude $
\langle g|g' \rangle$. This can be computed using any convenient
basis $\{ |i\rangle \}$ for the Hilbert space $\mathcal{H}$, as
$\langle g|g' \rangle = \sum_i \langle g|i \rangle \langle i| g'
\rangle$. One could use the basis we mentioned earlier $|\mu,m,\bar{m}
\rangle$ and compute the matrix elements of the operator matrix
$\hat{g}$; in principle this would allow one to deduce $\langle
\mu,m,\bar{m} |g \rangle$. It would be nice to perform such a
calculation, however instead we take a shortcut. Note that
\cite{fulop} has addressed the position representation starting from
the Klein-Gordon equation. In a position
representation we seek eigenfunctions of the current algebra which can
be represented by the differential operators,

\begin{eqnarray}
L_2 = i\hbar\frac{\partial}{\partial v}, \qquad R_2 =
-i\hbar\frac{\partial}{\partial u}, \\
L_{\pm} = i\hbar e^{\mp 2iv} \left( \coth 2\rho \frac{\partial}{\partial v}
- \cosech 2\rho  \frac{\partial}{\partial u} \pm i
\frac{\partial}{\partial \rho} \right), \\
R_{\pm} = i\hbar e^{\pm 2iu} \left( \coth 2\rho \frac{\partial}{\partial u}
- \cosech 2\rho  \frac{\partial}{\partial v} \mp i
\frac{\partial}{\partial \rho} \right).
\end{eqnarray}

It is straightforward to check that they satisfy the correct algebra

\begin{eqnarray}
[L_a,L_b]= 2\hbar i\epsilon_{ab}^{\phantom{ab}c} L_c, \qquad [R_a,R_b]=
2i\hbar\epsilon_{ab}^{\phantom{ab}c} R_c, \qquad [L_a,R_b]=0,
\end{eqnarray}

where $L_{\pm}= L_3 \pm iL_1$. The Casimirs $Q_L =
\eta^{ab}L_a L_b$ and $Q_R$, turn out to be

\begin{equation}
Q_L = Q_R = -\hbar^2 \Box,
\end{equation}

where $\Box$ is the Laplacian on $AdS_3$.
Thus in an irreducible representation the eigenfunctions of
$L_2,R_2,Q$ satisfy the Klein-Gordon equation with a mass that depends
on which expression for the Casimir is used. Therefore the propagator
$\langle g | g' \rangle$ will be the same function as the Wightman
function for the QFT with a suitable replacement for the mass
parameter. The Wightman function $G$ for a scalar
field satisfying $(\Box -\xi)\phi =0$, is a function of the invariant
distance  $\sigma(x,x')=\frac{1}{2} \eta^{(4)}_{\mu\nu}
(x-x')^{\mu}(x-x')^{\nu}$ ($x$ is the coordinate in the embedding
space $\mathbb{R}^{2,2}$ corresponding to the group element $g$) , and
is given by~\cite{satoh} \footnote{They provide the expression on the
universal cover of $AdS_3$, but actually it is the same for $AdS_3$,
since the expression is periodic in $t$.},
 
\begin{equation}
G(z)= \frac{1}{4\pi} (z^2-1)^{-1/2} [ z+ (z^2 -1)^{-1/2}]^{1-\lambda}
\end{equation}

where $z=1+ \sigma + i\epsilon \sgn[\sin(t-t')]$ and $\lambda= 1+
\sqrt{1+\xi}$ for $\xi \geq 0$ and $\xi=-1$, and $\lambda=1\pm\sqrt{1+\xi}$ for $-1<\xi<0$ . 
Thus we end up with an expression for $\langle g|g' \rangle$ which equals $G(z)$
with $\xi = \mu^2 -1$ for $Q=-p^2+\hbar^2$ and $\xi=\mu^2$ for $Q=-p^2$. It is
interesting to note that $\xi=\mu^2-1$ appears to correspond to some sort of
critical coupling since $\lambda = 1+\mu $ in this case. It should be noted
that the well-known Brietenlohner-Freedman bound for the Klein-Gordon
equation in $AdS_3$ is $\xi \geq -1$ (in these
conventions)~\cite{breitenlohner},~\cite{mal}; this is consistent with
$\xi=\mu^2-1$ which corresponds to the Casimir $Q=-p^2 +\hbar^2$. Hence it
appears that the quantum mechanics with the quantum correction to the
Casimir is consistent with the field theory, despite the fact that
the definition of mass is somewhat arbitrary.

\section{Conclusions}

We studied the motion of a free massive particle moving on the group manifold
$AdS_3$ both classically and quantum mechanically in a covariant
canonical formalism. We derived two different quantum theories
depending on whether the constraint that the motion be future directed is
imposed classically or quantum mechanically. The allowed values of the mass of
a particle quantum mechanically are quantized in either case.  Only
certain representations of the current algebra were allowed in the
Hilbert space, namely (most of) the discrete series. Interestingly in
the unconstrained theory  the continuous series in the exceptional
interval is  present leading to a small continuum of mass states;
these states have the property that they can flip between moving
forward or backward in time, and of course have no classical
counterpart.

A quantization in terms of ``chiral'' variables of the
theory was carried out, which amounts to determining something called
a braiding matrix, and this gave the same results as canonical
quantization of the unconstrained theory.

Upon comparison with QFT we found that one of the quantum theories
corresponds to a critical coupling (neither minimal or conformal),
which is consistent with the Breitenlohner-Freedman bound. It
would also be interesting to see the precise role of boundary
conditions, which must feature
since $AdS_3$ is not globally hyperbolic \footnote{These correspond to the two
different values of $\lambda$ for $-1<\xi<0$.}, in the quantum mechanics.
 
Possible extensions of this work include a similar treatment for
massless particles, a more systematic derivation of the position representation, and of course attacking the problem of a canonical
quantization of string theory in $AdS_3$.

\acknowledgments
I would like to thank Malcolm Perry for many discussions and useful
comments. This work was supported by EPSRC.

\appendix

\section{Some representation theory}

In this section we summarise the unitary irreducible representations
of $SL(2,\mathbb{R})$ and its universal covering group
$\widetilde{SL(2,\mathbb{R})}$, see~\cite{barg}~\cite{mal}. Both have the same algebras, namely
$sl(2,\mathbb{R})$. Let $\{t_a : a=1,2,3 \}$ be a basis for the
algebra satisfying $[t_a,t_b]=
\epsilon_{ab}^{\phantom{ab}c}t_c$, where $\epsilon_{ab}^{\phantom{ab}c} = \epsilon_{abd}\eta^{dc}$ and
$\eta$ is
the diagonal metric with $\eta_{11}=\eta_{22}=-\eta_{33}=1$. Consider a unitary representation
$R: t_a \mapsto -iJ_a$, so $J_a^{\dagger} = J_a$. Therefore we have,

\begin{equation}
[J_a,J_b]=i\epsilon_{ab}^{\phantom{ab}c} J_c.
\end{equation}
 
Define $J_{\pm}=J_1 \pm iJ_2$. The Casimir operator is $Q=\eta_{ab}J_a
J_b =J_1^2 +J_2^2 - J_3^2$. Now, lets examine the spectrum of $J_3$ in
an irreducible unitary representation; necessarily the Casimir will be
 diagonal.

\begin{eqnarray}
Q|j,m \rangle =-j(j+1)|j,m \rangle \\
J_3 |j,m \rangle = m|j,m \rangle
\end{eqnarray}

where $j\equiv -\frac{1}{2} - \sqrt{\frac{1}{4}-q}$ and $q$ is the
eigenvalue of $Q$. Since $J_a$ is Hermitian, $\{ |j,m\rangle \}$ can be
chosen to be an orthonormal basis for the carrier space to the
representation, also $m$ must be real. It is easy to show that,

\begin{equation}
J_{\pm}|j,m\rangle = \sqrt{m(m\pm 1) - j(j+1)} |j,m\pm1 \rangle
\end{equation} 

So far everything stated applies equally to $SL(2,\mathbb{R})$ and
$\widetilde{SL(2,\mathbb{R})}$. The differences arise from the
differences in the spectrum of $J_3$. By considering the
representation induced on the enveloping algebra it is easy to show
that $m$ is either integer or half-integer in the case of
$SL(2,\mathbb{R})$, whereas this restriction fails for
$\widetilde{SL(2,\mathbb{R})}$.
\begin{enumerate}
\item
The principal discrete representations correspond to highest and lowest weight
representations. More explicitly, $D^{+}_j$ is defined by $J_-|j,-j \rangle =
0$, and thus $m=-j,-j+1,...$ and $j<0$. Similarly $D^{-}_j$ is defined
by $J_+|j,j\rangle = 0$, and therefore $m=j,j-1,...$ and $j<0$. For
$\widetilde{SL(2,\mathbb{R})}$ $j$ is not restricted further, whereas
for $SL(2,\mathbb{R})$ $2j \in -\mathbb{N}$.
\item
The principal continuous representations correspond to no highest or lowest weight, so the
spectrum of $J_3$ is unbounded. Also, $j=-\frac{1}{2}+i\kappa$ and
$m=\alpha, \alpha \pm 1,...$, where $\kappa \in \mathbb{R}-\{0\}$ and $0\leq
\alpha <1$. These representations are labelled by $C^{\alpha}_j$, and
 $\forall \alpha \in [0,1)$ correspond to irreducible representations of $\widetilde{SL(2,\mathbb{R})}$,
and for $\alpha = 0,\frac{1}{2}$ gives two inequivalent irreducible representations of
$SL(2,\mathbb{R})$.
\item
There is another set of representations for which the spectrum of $J_3$
is unbounded. These are called the exceptional representations, for which
$-1< j<-\frac{1}{2} $ ($0<q<\frac{1}{4}$) and $m=\alpha, \alpha \pm 1,...$, and will be
denoted by $E^{\alpha}_j$. Note these are often also called
$C^{\alpha}_j$ or $C^{\alpha}_q$,
and there is no confusion with the  representations above since we
have different values of $j$. Here, $\alpha=0$ corresponds to
 representations of $SL(2,\mathbb{R})$.

\end{enumerate}

\section{Reparameterisation invariance at the quantum level}

Here we provide justification for why gauge fixing the Lagrangian
$\mathcal{L}=\frac{1}{e}\dot{x}^2 - m^2e$ by letting $e=1$ does not
spoil reparameterisation invariance quantum mechanically; see
also~\cite{pol} for a different argument. So to begin
we note that the Lagrangian is singular, leading to the primary
constraint $\phi_1=p_e \equiv \frac{\partial\mathcal{L}}{\partial \dot{e}}
\approx 0$. We want to also impose the constraint $\phi_2=e-1\approx
0$. If we look for secondary constraints we find $\phi_3=
\frac{1}{4}\pi^2 + m^2 \approx 0$, where $\pi_{\mu}$ is conjugate to
$x^{\mu}$, and no more occur. It is easy to see that $\phi_3$ is
first-class, and $\phi_1$, $\phi_2$ are second class. Now, we replace
Poisson brackets on the phase space with canonical variables
$(x,\pi,e,p_e)$, with the Dirac bracket defined by $\{ A,B \}_D=\{ A,B
\} - \{A,\phi_a \} C^{-1}_{ab} \{ \phi_b, B \}$ for any two
observables $A,B$, where $C_{ab}= \{ \phi_a, \phi_b \}$ and $a,b \in
\{1,2 \}$. Now, using the Dirac brackets instead of the Poisson
brackets, allows us to set $\phi_1$ and $\phi_2$ strongly to zero;
therefore $\{A,\phi_a \}=0$ for any observable $A$, since once the
second class constraints are set strongly to zero $A$ will only be a
function of $(x,\pi)$. This allows us to deduce that $\{A,B \}_D=\{A,B \}$. Hence the Lagrangian $\mathcal{L}= \dot{x}^2$ (the
constant $m^2$ isn't important and will be dropped) with
the constraint $\phi_3$ will give rise to the same phase space and
equations of motion as the manifestly reparameterisation invariant
Lagrangian. Also, the constraint is easy to implement quantum
mechanically giving rise to the BRST operator $Q=\phi_3 c$ which is
nilpotent (since $c$ is a ghost). The physical state conditions are
$Q|\psi\rangle =0$ and $b|\psi \rangle = 0$, where $b$ is the
antighost and $\{ b,c \}=1$; this implies $\phi_3 |\psi \rangle = 0$,
which becomes $p^2|\psi \rangle = m^2|\psi \rangle$.

\end{document}